\documentclass[PRB,reprint,amsmath,amssymb,superscriptaddress,longbibliography]{revtex4-1}
\usepackage{graphicx}
\usepackage{dcolumn}
\usepackage{bm}
\usepackage{hyperref}
\usepackage{breakurl}
\usepackage{multirow}
\usepackage{enumitem}
\usepackage{color}

\begin{document}

\title{Quantum transport in a compensated semimetal W$_{2}$As$_{3}$ with nontrivial $Z_{2}$ indices }

\author{Yupeng Li}
      \thanks{Equal contributions}
      \affiliation{Department of Physics, Zhejiang University, Hangzhou 310027, People's Republic of China}
      \affiliation{State Key Laboratory of Silicon Materials, Zhejiang University, Hangzhou 310027, People's Republic of China}

\author{Chenchao Xu}
      \thanks{Equal contributions}
      \affiliation{Department of Physics, Zhejiang University, Hangzhou 310027, People's Republic of China}

\author{Mingsong Shen}
      \affiliation{Wuhan National High Magnetic Field Center, School of Physics, Huazhong University of Science and Technology, Wuhan 430074, People's Republic of China}

\author{Jinhua Wang}
      \affiliation{Wuhan National High Magnetic Field Center, School of Physics, Huazhong University of Science and Technology, Wuhan 430074, People's Republic of China}

\author{Xiaohui Yang}
      \affiliation{Department of Physics, Zhejiang University, Hangzhou 310027, People's Republic of China}

\author{Xiaojun Yang}
      \affiliation{School of Physics and Optoelectronics, Xiangtan University, Xiangtan 411105, People's Republic of China}

\author{Zengwei Zhu}
      \affiliation{Wuhan National High Magnetic Field Center, School of Physics, Huazhong University of Science and Technology, Wuhan 430074, People's Republic of China}

\author{Chao Cao}
      \email{ccao@hznu.edu.cn}
      \affiliation{Department of Physics, Hangzhou Normal University, Hangzhou 310036, People's Republic of China}

\author{Zhu-An Xu}
      \email{zhuan@zju.edu.cn}
      \affiliation{Department of Physics, Zhejiang University, Hangzhou 310027, People's Republic of China}
      \affiliation{State Key Laboratory of Silicon Materials, Zhejiang University, Hangzhou 310027, People's Republic of China}
      \affiliation{Collaborative Innovation Centre of Advanced Microstructures, Nanjing 210093, People's Republic of China}

\date{\today}

\begin{abstract}

We report a topological semimetal W$_{2}$As$_{3}$ with a space
group C2/m. Based on the first-principles calculations, band
crossings are partially gapped when spin-orbit coupling is
included. The $Z_{2}$ indices at the electron filling are [1;111],
characterizing a strong topological insulator and topological surface states.
From the magnetotransport measurements, nearly quadratic field dependence
of magnetoresistance (MR) ($B\parallel[200]$) at 3 K indicates an
electron-hole compensated compound whose longitudinal MR reaches
11500\% at 3 K and 15 T. In addition, multiband features are
detected from the high-magnetic-field Shubnikov-de Haas (SdH)
oscillation, Hall resistivity, and band calculations. A nontrivial
$\pi$ Berry's phase is obtained, suggesting the topological
feature of this material. A two-band model can fit well the
conductivity and Hall coefficient. Our experiments manifest
that the transport properties of W$_{2}$As$_{3}$ are in good
agreement with the theoretical calculations.

\end{abstract}

\maketitle

\section*{Introduction}

The research of new topological phases such as topological
insulator (TI), topological superconductor, and topological
semimetal (TSM) is a hot spot in recent years in condensed matter
physics. Among them, topological semimetals are widely studied
because there are many types, including Dirac semimetals
\cite{Na3Bi_WangZJ_PRB12,Cd3As2_WangZJ_PRB13}, Weyl semimetals
\cite{WeylSemi_WHM_PRX}, nodal-line semimetals
\cite{PbTaSe2_BianG_NatC,Cu3PdN_DNLS_PRL2015_WengHM,Cu3NPd_YKKim_PRL2015,ZrSiO_XuQS_PRB,SrIrO3_FangC_PRB2015,CaAgX_AYk_JPCJ2016},
and semimetals with triply degenerate nodal points
\cite{triply_Bradlyn_science16,triply_WHM_PRB16,triply_BQLv_nature17}.
In addition, they are closely related to other topological phases 
and are thus believed to be intermediate states of topological
phase transitions \cite{Na3Bi_LiuZK_science14,Iridates_WanXG_PRB2011}. 
These topological semimetals usually exhibit large magnetoresistance and
topological surface states. Apart from these topological phases,
another kind of topological materials are Z$_{2}$ topological
metals \cite{LaBiARPES_NayakJ_ncom17}, which are characterized by
non-trivial Z$_{2}$ topological invariants, topological surface
states, and lack of a bulk energy gap, such as the La$X$ ($X$ = P,
As, Sb, Bi) \cite{LaXTI_LinHsin_arXiv} family. Among this family,
LaBi is the only one with topological nontrivial band dispersion
proved by angle-resolved photoemission spectroscopy (ARPES)
measurements \cite{LaBiARPES_NayakJ_ncom17}.

Recently, transition-metal dipnictides MPn$_{2}$ (M =  Nb or Ta;
Pn = As or Sb) with C2/m structure have attracted great attention
because of extremely large magnetoresistance (XMR), negative
magnetoresistance and other interesting properties
\cite{TaSb2_LYK_PRB,NbAs2family_YupengLi_arXiv,NbAs2_ShenB_PRB,TaAs2_FangZ_APL14,TaAs2_SciR_LuoYK,TaAs2_ShuangJ_PRB,TaAs2_XiaTL_PRB}.
According to the theoretical calculations, type-II Weyl points
could be induced by magnetic field in MPn$_{2}$
\cite{NbAs2Weyl_GreschD_NJP17}. Moreover, a superconducting
transition has been observed in NbAs$_{2}$ at high
pressure\cite{NbAs2SC_Liyp}, and the maximal $T_{c}$ is 2.63 K
under 12.8 GPa. What is very interesting is that the C2/m
structure remains up to 30 GPa, and thus the topological phase
could exist in the superconducting state and it may be a candidate
of topological superconductors. MoAs$_{2}$ is another compound
with C2/m structure possessing quadratic XMR
\cite{MoAs2_LiYL_SR17,MoAs2ARPES_LuoR_PRB17}, whose XMR may
originate from the open-orbit topology
\cite{MoAs2ARPES_LuoR_PRB17} instead of the electron-hole
compensation in MPn$_{2}$.

\begin{figure*}[!thb]
\begin{center}
\includegraphics[width=7in]{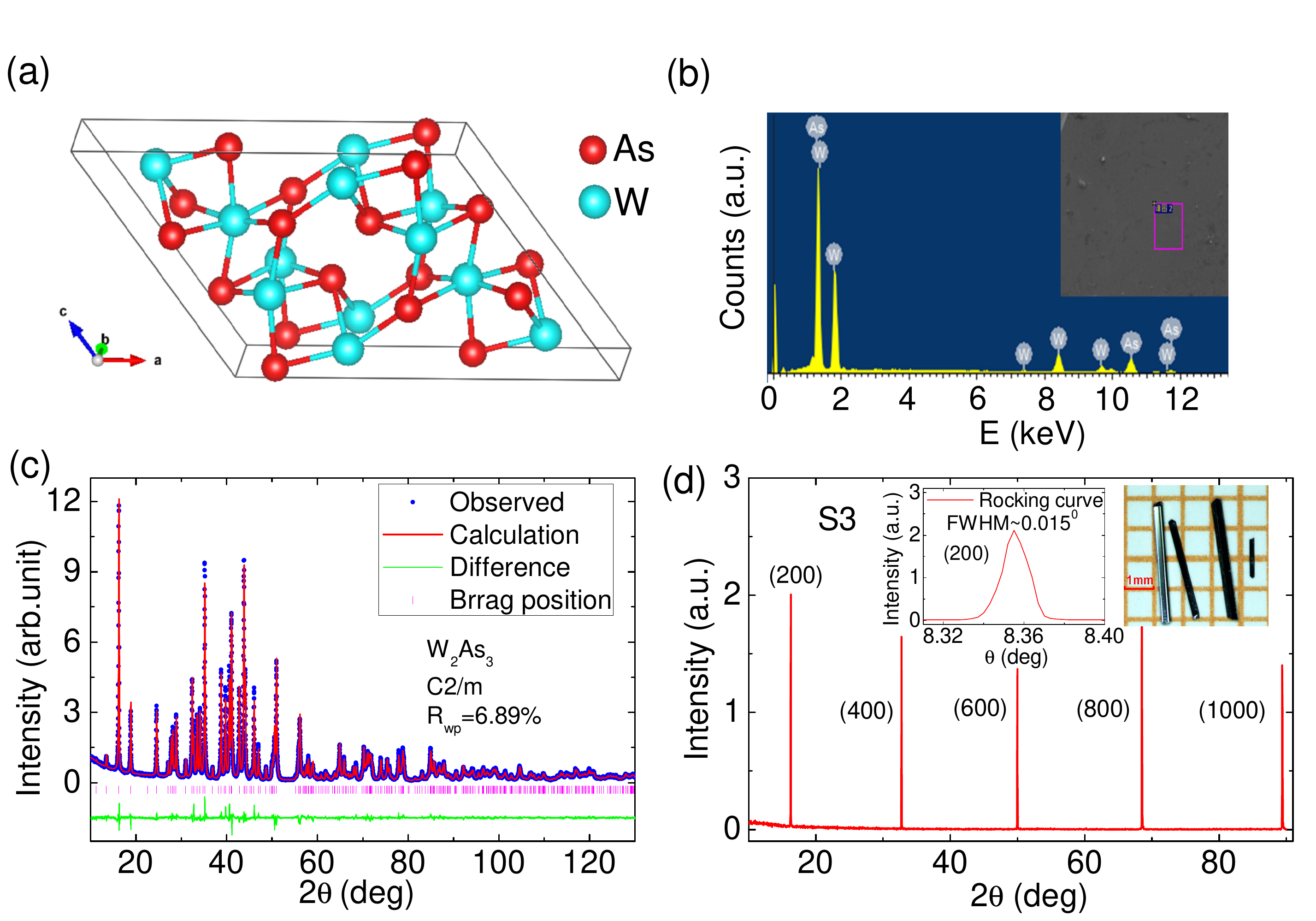}
\end{center}
\caption{\label{Fig1} (a) Crystalline structure of W$_{2}$As$_{3}$.
(b) EDS spectrum of W$_{2}$As$_{3}$ and the inset is a scanning electron
microscopy figure. (c) Powder XRD pattern of polycrystalline
W$_{2}$As$_{3}$ and Rietveld analysis profiles. (d) XRD pattern of
high-quality single crystal showing sharp diffraction peaks of the
(200) plane. The left inset is the rocking curve of the (200) peak,
and the right inset is a photograph of W$_{2}$As$_{3}$ crystals. }
\end{figure*}

These interesting properties in the compounds with C2/m structure
have attracted much research attention. Here we report on a 
topological material W$_{2}$As$_{3}$ with C2/m structure and
nontrivial $Z_{2}$ indices [1:111] which belongs to a strong TI
family
\cite{Z2calculation_FuL_PRB07,CaAgX_AYk_JPCJ2016,Cu3PdN_DNLS_PRL2015_WengHM,Bi1-xSbx_TJCY_PRB2015,TI_RMP_Hasan}.
Large MR$=[R(B)-R(0)]/R(0)$ of approximately 11 500\% at 3 K and 15
T is observed in this electron-hole compensated system. Evident
quantum oscillations have been observed by using a high pulse
magnetic field, from which nine intrinsic frequencies are obtained
from the fast Fourier transform (FFT) spectrum, and a nontrivial
$\pi$ Berry's phase can be detected from $F_{h3}$, $F_{h4}$, and
$F_{h5}$. The multiband character is also revealed from both band
calculations as well as Hall resistivity measurements. Negative
Hall resistivity indicates that this compound is an electron-dominated semimetal. 
A simple two-band model is proposed to
explain the temperature dependence of the Hall coefficient, with which
the Hall conductivity and longitudinal conductivity can be well
fitted. The Hall coefficient obtained from the fitting data is in
agreement with the experimental value of $R_{H}$. All of these
illustrate relatively consistent transport behavior and the
reliable two-band fitting.

\begin{figure*}[!thb]
\begin{center}
\includegraphics[width=7in]{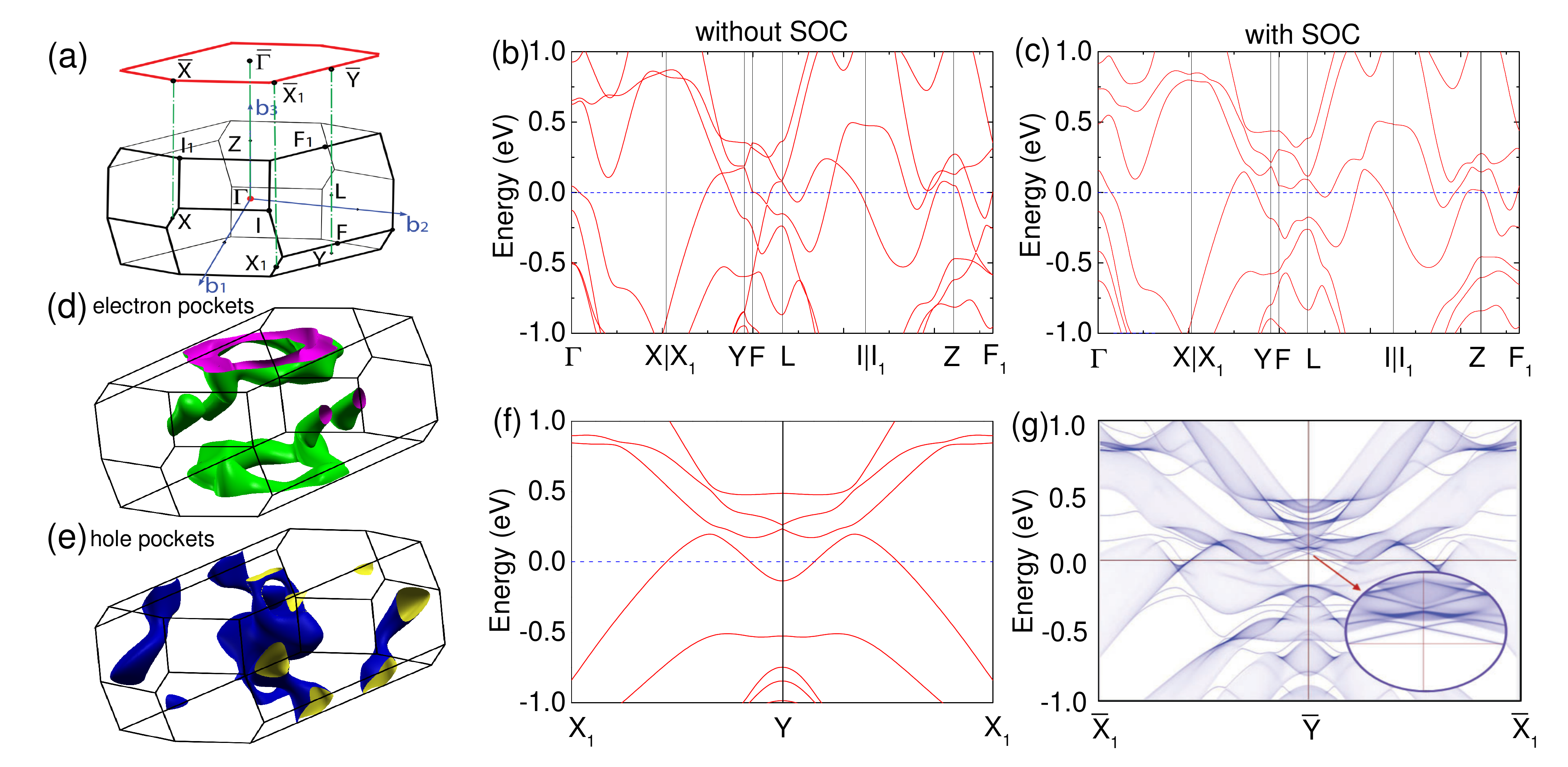}
\end{center}
\caption{\label{Fig1} (a) A schematic diagram of primitive
Brillouin zone and the corresponding projection of the (001)
surface Brillouin zone. (b) Band structure of W$_{2}$As$_{3}$
without SOC. (c) The band structure with SOC. (d, e) The
3D Fermi surfaces of electron pockets (green ones) and hole
pockets (blue ones) with SOC, respectively. (f) Expanded view
of bulk band structure with SOC along $Y-X_{1}$. (g) Calculated
(001) surface band structure along the
$\overline{X}_{1}-\overline{Y}-\overline{X}_{1} $ line, and the
inset shows the topological surface state. }
\end{figure*}

\section*{Experiment}

Single crystals of W$_{2}$As$_{3}$ were synthesized by an
iodine-vapour transport method. Firstly, polycrystalline
W$_{2}$As$_{3}$ was prepared by heating the mixture of W powder
and As powder with a stoichiometric ratio of 2:3 at 1173 K for 2
days. Then iodine with a concentration of 10mg/cm$^{3}$ was mixed
with the reground polycrystalline W$_{2}$As$_{3}$, and they were
sealed in an ampoule to grow single crystals for 7 days in a
two-zone furnace, where the temperature gradient was set as 1323-1223 K 
over a distance of 16 cm. All the experimental processes
were carried out in a glove box filled with pure Ar, except the
heating processes.

\begin{figure*}[!thb]
\begin{center}
\includegraphics[width=7in]{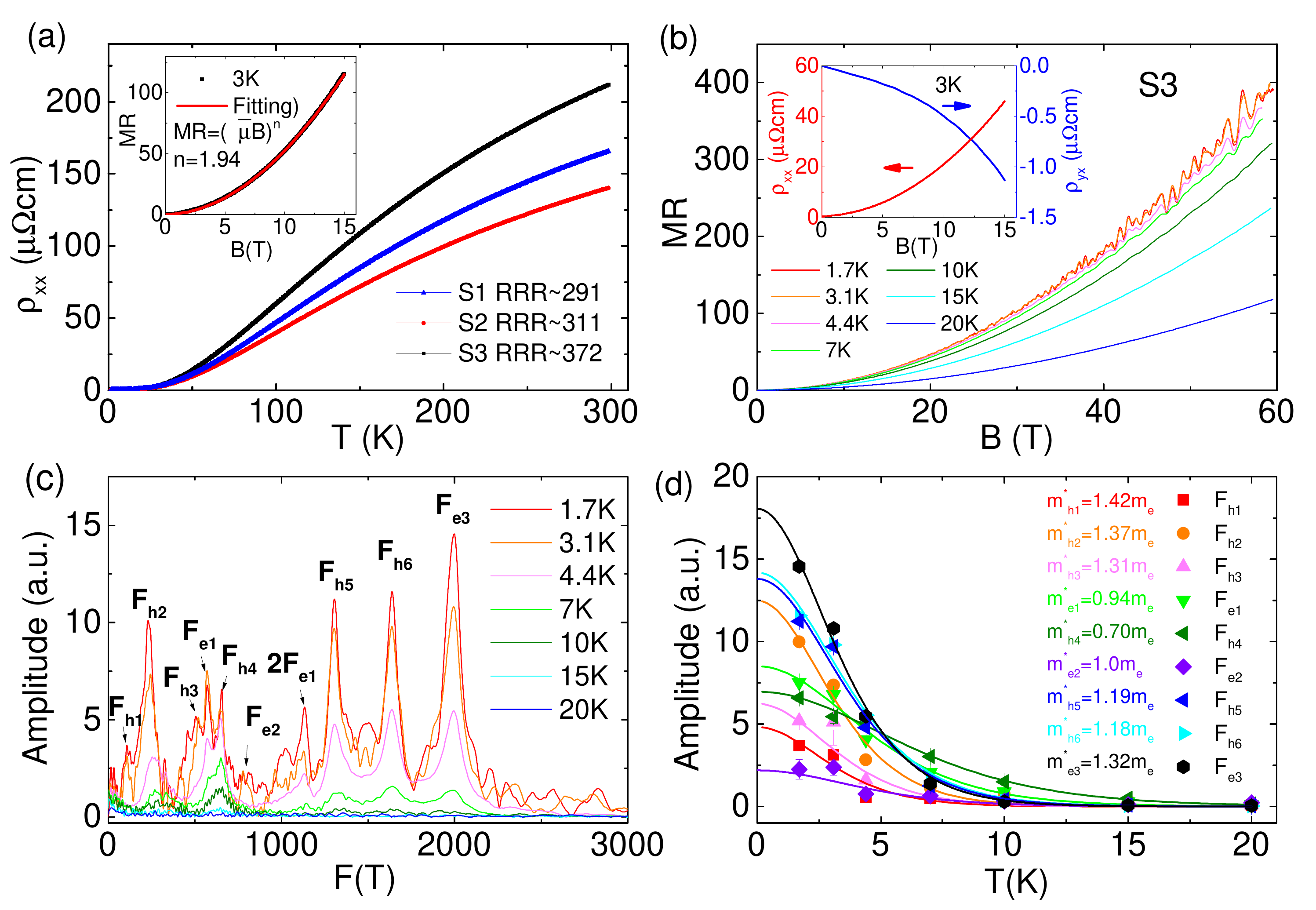}
\end{center}
\caption{\label{Fig3} (a) Temperature-dependent resistivity of
several samples. The inset is a fitting result with $MR=(\bar{\mu}
B)^{n}$. (b) Visible quantum oscillations under high magnetic
field. The inset displays $\rho_{xx}$ measured up to 15 T which is
much larger than $\rho_{yx}$. (c) FFT curves suggesting the nine
independent intrinsic frequencies. (d) Fitting the SdH amplitudes
by the LK formula. The obtained effective masses of the nine FFT
frequencies are shown.}
\end{figure*}

X-ray diffraction (XRD) data were collected by a PANalytical x-ray
diffractometer (Empyrean) with a Cu K$_{\alpha1}$ radiation. We used
energy-dispersive x-ray spectroscopy (EDS) to analyze the composition
ratio of W and As element. Longitudinal resistivity and Hall resistivity
was measured by a standard six-probe technique. Transport measurements
below 15 T were performed on an Oxford-15T cryostat, and a pulsed magnetic
field up to 60 T was employed to obtain apparent quantum oscillations
in the Wuhan National High Magnetic Field Center (WHMFC - Wuhan).

The first-principles calculations were performed with Vienna Abinitio
Simulation Package (VASP) \cite{VASP_Kresse_PRB93,VASP_Kresse_PRB96}.
A plane-wave basis up to 400 eV was employed in the calculations.
Throughout the calculation, the atom relaxation was performed with the
Perdew-Burke-Ernzerhof (PBE) exchange correlation functional, and the 
electronic band structures, Fermi surfaces, and surface states were obtained with
the modified Becke-Johnson (mBJ) method \cite{mBJ_Tran_PRL102}. A
$\Gamma$-centred 15$\times$15$\times$6 Monkhorst-Pack \cite{MPscheme_Monkhorst_PRB76}
k-point mesh was chosen to sample the Brillouin zone for the calculation.
The crystal structure was fully optimized until the force on each atom was
less than 1 meV/\AA\ and internal stress less than 0.1 kbar. The calculated
lattice constants as well as the atomic coordinates were within a 5\% errorbar
compared to the refinement results of XRD data by the software Rietan-FP \cite{Rietan2007}.
The topological indices $Z_{2}$ were calculated by the method of
parity check\cite{Z2calculation_FuL_PRB07}. Using the maximally localized
Wannier function method\cite{MLWF}, the Fermi surfaces were obtained with
a tight-binding Hamiltonian fitted from density functional theory (DFT) bands. 
The surface states were calculated using surface a Green's function \cite{surfacestate_1985JPF}.

\section*{Results}

The crystal structure (a unit cell) of W$_{2}$As$_{3}$ is
presented in Fig. 1(a) with the monoclinic C2/m (space group,
no. 12), which has the same space group with NbAs$_{2}$ \cite{NbAs2family_YupengLi_arXiv,NbAs2familiyculc_caochao_PRB,TaSb2_LYK_PRB,MoAs2ARPES_LuoR_PRB17,NbAs2_ShenB_PRB,TaAs2_ShuangJ_PRB}.
EDS data in Fig. 1(b) give the chemical component ratio
W:As = 2:2.85, consistent with the nominal ratio of 2:3 within the
experiment error. Rietveld refinement of powder XRD is fairly
reliable with R$_{wp}$=6.89\%, as shown in Fig. 1(c). The refined
lattice parameters at room temperature are $a$ = 13.322 {\AA}, $b$
= 3.277 {\AA}, $c$ = 9.593 {\AA} and $\beta$ = 124.704$^\circ$,
which are almost the same as in the previous
report \cite{W2As3_TJB_CJC1965}. Subsequently, a single-crystal XRD
pattern of sample S3 is shown in Fig. 1(d), and the rocking
curve in the left inset indicates a high-quality sample (S3) with
a small FWHM = 0.015$^{\circ}$ \cite{Xinghui_JPCM2011}. The obtained
single-crystal samples are usually needle-shaped, as shown in the
inset of Fig. 1(d).

\begin{table}
\tabcolsep 0pt \caption{\label{parity} Parities of bands at time-reversal
invariant momenta (TRIM). $\Pi_n$ is the multiplication of the parities for
bands 1 to $n$. The highest occupied band at each TRIM is indicated with $^\circ$.
The corresponding $Z_{2}$ classification is [1,111].}
\vspace*{-12pt}
\begin{center}
\def\temptablewidth{1.0\columnwidth}
{\rule{\temptablewidth}{1pt}}
\begin{tabular*}{\temptablewidth}{@{\extracolsep{\fill}}ccccc}
$\quad$ $\quad$ TRIM    $\quad$ $\quad$  &$\prod_{26}$ $\quad$ &$\prod_{27}$ $\quad$  &$\prod_{28}$     $\quad$ &$\prod_{29}$ $\quad$\\ \hline
(0,0,0)             & $-^\circ$ \ \ \   &  $-$ $\quad$ \        &  $+$ $\quad$&  $-$   $\quad$ \\
($\pi$,0,0)         &$+$   $\quad$     & $-^\circ$ $\quad$   &  $-$ $\quad$  & $+$  $\quad$\\
(0,$\pi$,0)         & $+$   $\quad$     & $-^\circ$ $\quad$  &  $-$  $\quad$ & $+$   $\quad$\\
($\pi$,$\pi$,0)     & $-$   $\quad$     & $+^\circ$ $\quad$  &  $-$  $\quad$ & $-$   $\quad$\\
(0,0,$\pi$)         & $-^\circ$ \ \ \    & $+$     $\quad$ \    &  $+$  $\quad$ & $+$   $\quad$\\
($\pi$,0,$\pi$)     & $-$   $\quad$     & $+^\circ$ $\quad$   &  $+$ $\quad$  & $+$  $\quad$\\
(0,$\pi$,$\pi$)     & $-$   $\quad$    & $+^\circ$  $\quad$  &  $+$  $\quad$ & $+$   $\quad$\\
($\pi$,$\pi$,$\pi$) & $-$   $\quad$    & $+^\circ$  $\quad$  &  $-$  $\quad$ & $-$   $\quad$\\
\end{tabular*}
{\rule{\temptablewidth}{1pt}}
\end{center}
\vspace*{-18pt}
\end{table}

\begin{figure}[!thb]
\begin{center}
\includegraphics[width=3.5in]{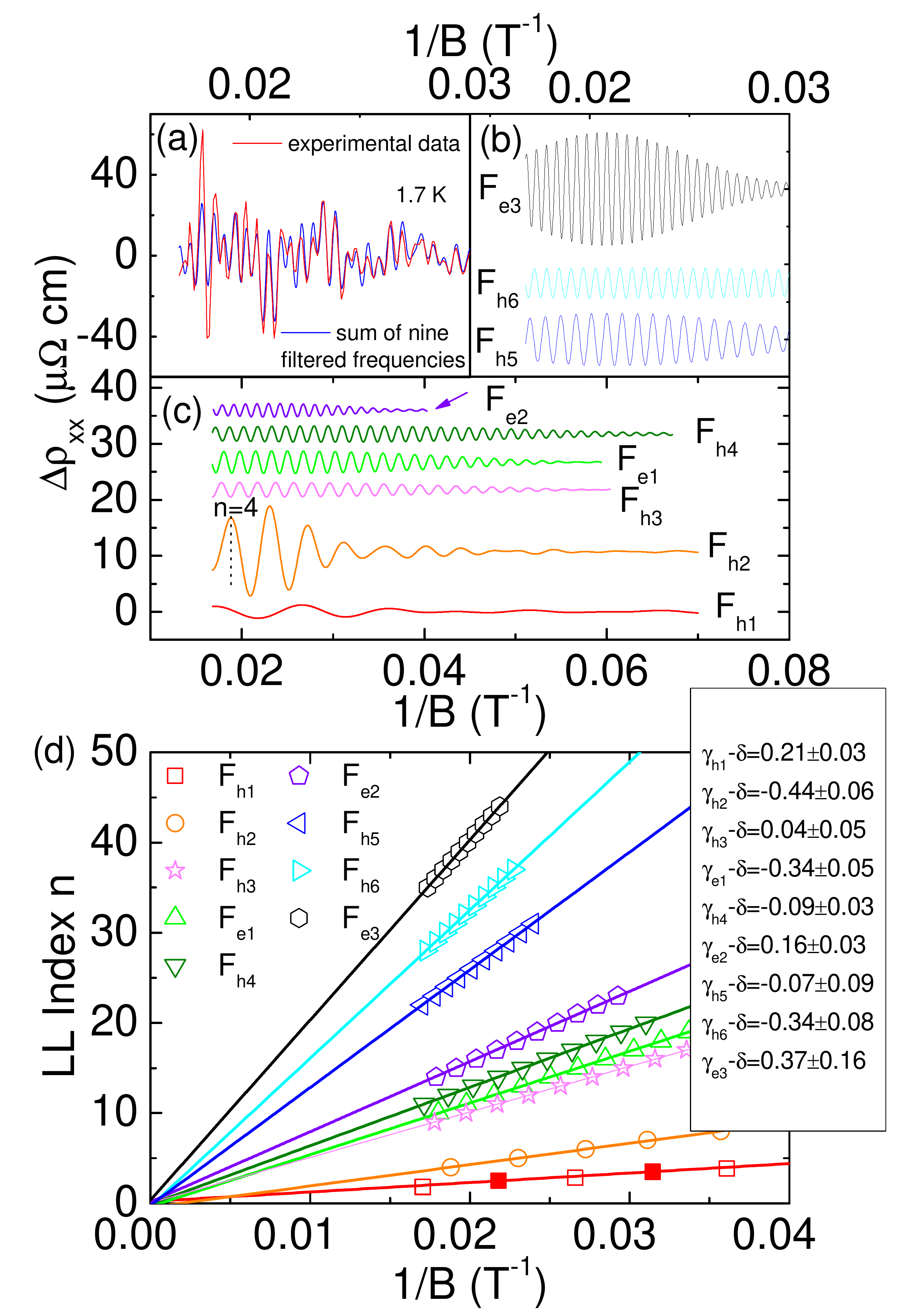}
\end{center}
\caption{\label{Fig4} (a) Oscillatory part of the resistivity (
$\Delta\rho_{xx}$) as a function of 1/B at 1.7 K (the red line is
experimental data). $\Delta\rho_{xx}$ is obtained after
subtracting the background of $\rho_{xx}$. The blue line is the
sum of nine filtered frequencies. (b, c) The nine
filtered oscillatory parts of $\Delta\rho_{xx}$. The Landau-level
index is obtained from the peak position of the oscillation
component, for example, one of the oscillatory peaks about $F_{h2}$
is $n$ = 4 in (c). (d) Landau-level index plot of the nine
frequencies. The intercepts are between $-$1/8 and 1/8 for
$F_{h3}$, $F_{h4}$, and $F_{h5}$, respectively. Open circles
indicate the integer Landau-level index from the peaks of
high-frequency oscillatory components of $\Delta\rho_{xx}$, and
closed circles denote the half-integer index ($\Delta\rho_{xx}$
valley). }
\end{figure}

The band structures of W$_{2}$As$_{3}$ without and with spin-orbit
coupling (SOC) are shown in Fig.2, respectively. Like the
TaSb$_{2}$ compound\cite{NbAs2familiyculc_caochao_PRB} , there is a 
one-electron band and a one-hole band crossing the Fermi level,
indicating the two-band feature of this system. W$_{2}$As$_{3}$,
however, has large Fermi surfaces (FS) [Figs. 2(d) and (e)], and three
band-crossing features can be identified near the Fermi level.
They are along the X$_{1}$Y, FL and LI directions, respectively, which
all resemble the transition metal dipnictides XPn$_{2}$ (X = Ta, Nb;
Pn = P, As, Sb) \cite{NbAs2familiyculc_caochao_PRB}. Once the SOC is
included, all the band crossings are gapped, leaving these two
bands separated near the Fermi level $E_{f}$. Therefore, this
system can be adiabatically deformed to an insulator without
closing or opening gaps, thus allowing us to define the Z$_{2}$
index with the parity at the time-reversal invariant momenta
(TRIM) multiplied up to the highest valence band, similar to the
cases of LaBi \cite{LaBiARPES_NayakJ_ncom17} and CeSb
\cite{CeSb_GuoCY_NPJqm17}. As both the highest valence band (band
27) and lowest conduction band (band 28) are crossing the Fermi
level, we list the band parities up to four bands $\Pi_{26}$,
$\Pi_{27}$, $\Pi_{28}$, and $\Pi_{29}$ in Table \ref{parity}, where
$\Pi_n$ is the multiplication of the parities for bands 1 to $n$.
There are 54 valence electrons in the primitive cell,
corresponding to 27 filled bands. The product of parities at all
these TRIMs up to the 27th band is -1, suggesting a strong topological
property of this compound. In addition, we also performed the PBE
calculations and obtained the same topological properties as the
mBJ calculations, in contrast to the divergence between the PBE
and mBJ calculations in LnPn (Ln = Ce, Pr, Gd, Sm, Yb; Pn = Sb, Bi)
\cite{LnPn_DuanX_arxiv18} and LaSb
\cite{LaSbPBEandmBJ_GuoPJ_PRB16}, in which the PBE calculations
underestimate the band gap between conduction band and valance
band. Compared with the bulk band structure along $X_{1}-Y$ in
Fig. 2(f), we calculate the (001) surface states along
$\overline{X}_{1}-\overline{Y}-\overline{X}_{1} $ as shown in
Fig. 2(g). Although there are strong bulk states due to the large
hole and electron pockets in Fig. 2(g), the Dirac type of surface
states can be still observed at the $\overline{Y}$ point in the inset
of Fig. 2(g).


\begin{table}
\tabcolsep 0pt \caption{\label{Tab.FFT} Physical parameters of nine
extremal orbits are listed when magnetic field is along the (200)
facet of W$_{2}$As$_{3}$. $m_{0}$ is the static mass of the electron.
DFT results here are the SdH frequencies from calculated band energies \cite{FFTcal_Rourke_CPC12}.  }
\vspace*{-12pt}
\begin{center}
\def\temptablewidth{1.0\columnwidth}
{\rule{\temptablewidth}{1pt}}
\begin{tabular*}{\temptablewidth}{@{\extracolsep{\fill}}ccccccc}
FS        &SdH(T)  &$m^{*}$($m_{0}$)      &DFT(T)     \\ \hline
$F_{h1}$    &106     &1.42                    &58        \\
$F_{h2}$    &232     &1.37                    &140       \\
$F_{h3}$    &505     &1.31                    &181       \\
$F_{e1}$    &574     &0.94                    &577       \\
$F_{h4}$    &653     &0.70                    &605       \\
$F_{e2}$    &778     &1.00                    &623       \\
$F_{h5}$    &1305    &1.19                    &1262      \\
$F_{h6}$    &1640    &1.18                    &1460      \\
$F_{e3}$    &1997    &1.32                    &1534       \\
\end{tabular*}
{\rule{\temptablewidth}{1pt}}
\vspace*{-18pt}
\end{center}
\end{table}

We now turn to the transport properties of W$_{2}$As$_{3}$. The
temperature-dependent $\rho_{xx}$ of different samples S1, S2, and
S3 all exhibit a typical metallic behavior, and the corresponding
residual resistance ratio [RRR =
$\rho_{xx}(300K)/\rho_{xx}(1.5K)$] is 291, 311, and 372,
respectively. Therefore, the sample S3 is chosen to perform further
studies. The giant MR of about 11 500\% at 15 T and 3 K is
displayed in the inset of Fig. 3(a) without obvious quantum
oscillations. The MR can be well fitted by $MR=(\bar{\mu}B)^{n}$
with $n=1.94$, indicating good electron-hole
compensation \cite{WTe2_MNA_Nature14,WTe2_ZZW_PRL}. The geometric
mean of the mobilities is $\bar{\mu}=\sqrt{\mu_{e}\mu_{h}}=0.72
T^{-1}$ when $n=2$ is used.

\begin{figure*}[!thb]
\begin{center}
\includegraphics[width=7in]{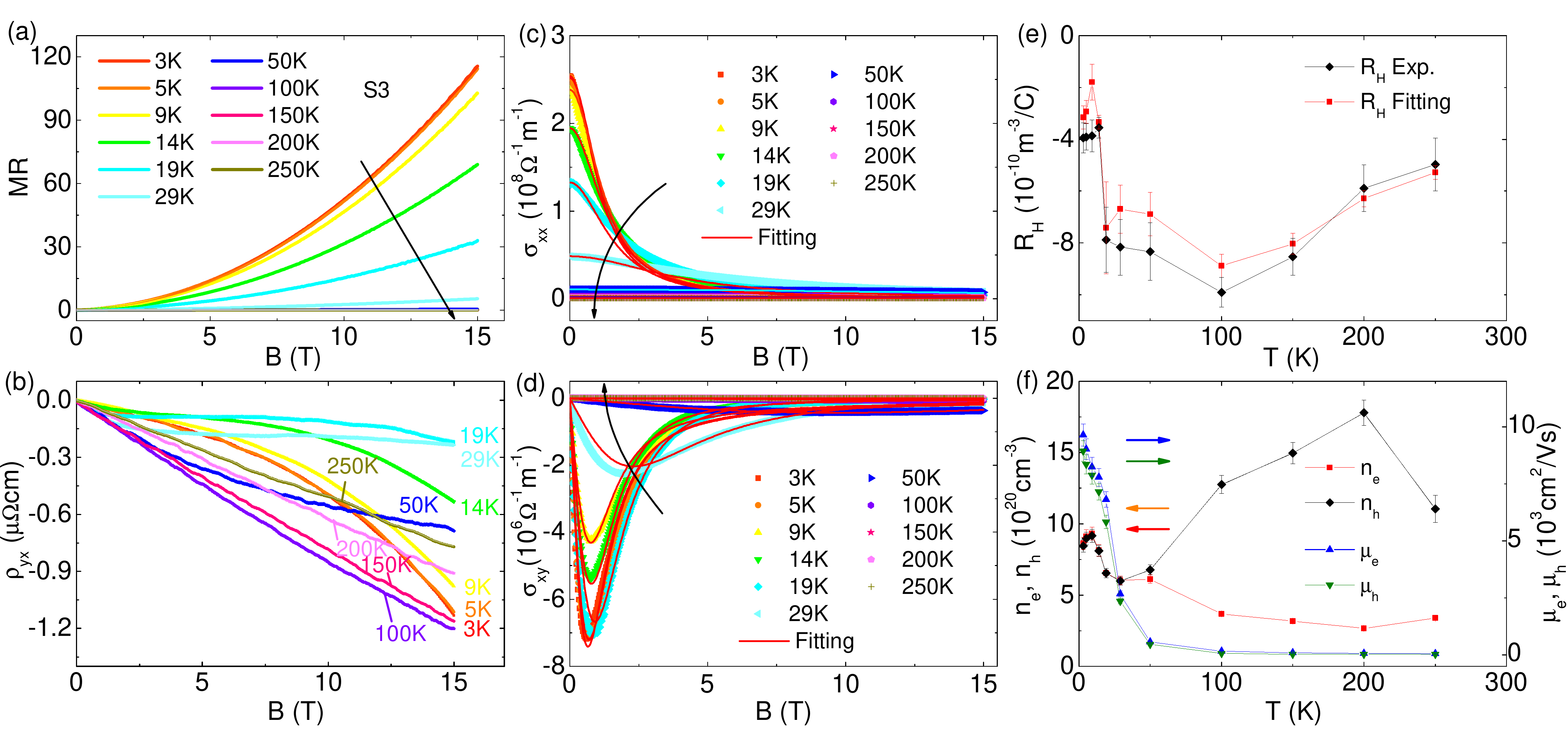}
\end{center}
\caption{\label{Fig5} (a) Magnetoresistance at various
temperatures; (b) Hall resistivity vs magnetic field; (c)
conductivity $\sigma_{xx}$ vs magnetic field; (d) Hall
conductivity $\sigma_{xy}$ vs magnetic field; and (e) temperature
dependence of Hall coefficient $R_{H}$. $R_{H}$ (solid square)
obtained from the two-band model fitting is compared with the
experimental data of $R_{H}$ (diamond). (f) Temperature dependence
of $n_{e}$, $n_{h}$, $\mu_{e}$, and $\mu_{h}$ obtained in the
two-band model fitting. }
\end{figure*}

In order to extract more information about the Fermi surface of
W$_{2}$As$_{3}$, a pulsed magnetic field experiment up to 60 T is
performed. In Fig. 3(b), large quantum oscillations in resistance
become visible at various temperatures, and no sign of saturation
in MR is detected up to 60 T. The inset of Fig. 3(b) shows
$|\rho_{yx}| \ll |\rho_{xx}|$, so we can use $\rho_{xx}$ to
analyze the quantum oscillation data. After removing the
background, nine distinct frequencies of extremal orbits are
observed from a complicated FFT spectrum in Fig. 3(c). The
calculated frequencies are in good agreement with the experimental
observations, considering the error between experimental results
and theoretical calculations \cite{FFTcal_Rourke_CPC12},
as shown in Table \ref{Tab.FFT}. Using the
Lifshitz-Kosevich (LK) formula,
\begin{equation}\label{LK}
\Delta R_{xx} \propto R_{T}\times R_{D} \times cos\left[2\pi(\frac{F}{B}+\gamma-\delta)\right],
\end{equation}
where $R_{T}=(2\pi^{2}k_{B}T/\beta)/sinh(2\pi^{2}k_{B}T/\beta)$,
$\beta=e\bar{B}\hbar/m^{*}$, $k_{B}$ is Boltzmann constant and $\bar{B}$ is the average field value \cite{WTe2averageB_RD_PRB,LaBi_Sunss_NJP}.
The obtained effective mass $m^{*}$ of each pocket is also
indicated in Fig. 3(d), where the smallest $m^{*}_{h4}=0.70m_{0}$
and largest $m^{*}_{h1}=1.42m_{0}$. Due to the complicated Fermi
surface, as seen in Figs.2(d) and 2(e), nine filtered oscillatory parts in Figs. 4(b) and 4(c)
are detected by the decomposition of $\Delta\rho_{xx}$ in Fig. 4(a), which is
obtained by subtracting the background of $\rho_{xx}$ at 1.7 K. Although both
the mussy frequencies and harmonic frequencies are eliminated, the sum of
nine main frequencies (blue line) matches the experimental data (red line) very well in Fig. 4(a), both in amplitudes and phases.
Therefore, we assign the Landau level by the peak
position of the oscillation component of each frequency in Figs. 4(b) and 4(c) \cite{ZrSiS_HuJ_PRB17,ZrSiS_MatusiakM_NComms17,QOBiSb_TaskinAA_PRB09}. In the Lifshitz-Onsager (LO)
formula,
\begin{equation}\label{LO}
A_{n}\frac{\hbar}{eB}=2\pi(n+\gamma-\delta),
\end{equation}
where $A_{n}$ is the FS cross section area of the Landau level (LL) $n$.
$\gamma-\delta=\frac{1}{2}-\frac{\phi_{B}}{2\pi}-\delta$ is the
phase factor, where $\phi_{B}$ is the Berry phase and $\delta$ is
a phase shift induced by dimensionality [$\delta=0$ for two dimensions (2D), or
$\delta=\pm\frac{1}{8}$ for three dimensions (3D)]. The Landau-level index is fitted by the 
LO formula in Fig. 4(d), and we can obtain the intercepts of nine
frequencies. Furthermore, $|\gamma-\delta|$ of $F_{h3}$, $F_{h4}$, 
and $F_{h5}$ are 0.04, 0.09 and 0.07, respectively, which are
all in the range between 0 and 1/8, exhibiting a nontrivial
Berry's phase of
$\pi$ \cite{Berryphase1_NKS_nature,Berryphase2_ZYB_nature}.
Theoretically, when a singularity in the energy band is enclosed
by the cyclotron contour under magnetic field, we could detect a
nontrivial $\pi$ Berry's phase from Shubnikov-de Haas (SdH)
oscillation \cite{BerryP_MGP_PRL1999,Cd3As2Angular_XiangZJ_PRL}. It
is worth noting that the observation of a $\pi$ Berry phase from
SdH oscillation is affected by the magnetic field directions and
the cross sections of Fermi surfaces; thus some nontrivial Fermi
surfaces may not yield exact a $\pi$ Berry phase in SdH
measurements \cite{BerryP_MGP_PRL1999,BerrtP_LuHZ_18,Cd3As2Angular_XiangZJ_PRL}.
For the materials with a complex Fermi surface, some unexpected
frequencies, such as low frequencies and harmonic frequencies,
would affect the filtered main frequency more or less. For example,
the harmonic frequency 2F$_{h4}$ (1306 T) may be mixed into the filtered
F$_{h5}$ (1305 T) and thus modulates the oscillation data of F$_{h5}$ in Fig. 4(b),
which could affect the precise determination of the Berry phase of F$_{h5}$ in Fig. 4(d).

The measurements of $\rho_{xx}(B)$ and $\rho_{yx}(B)$ provide
further insight on the transport properties, as seen in Fig. 5(a) and 
5(b). The negative $\rho_{yx}(B)$ shown in Fig. 5(b) indicates
that the electron-type carriers play a dominant role in transport
properties, and the nonlinear magnetic field dependence implies
the multi-band feature. Figures. 5(c) and 5(d) display that the curves
of conductivity
$\sigma_{xx}=\rho_{xx}/(\rho_{xx}^{2}+\rho_{yx}^{2})$ and
$\sigma_{xy}=\rho_{yx}/(\rho_{xx}^{2}+\rho_{yx}^{2})$ can be well
fitted by the two-band model:
\begin{gather}\label{twoband}
\sigma_{xx}=e\left[\frac{n_{h}\mu_{h}}{1+(\mu_{h}B)^2}+\frac{n_{e}\mu_{e}}{1+(\mu_{e}B)^2}\right],  \\
\sigma_{xy}=eB\left[\frac{n_{h}\mu_{h}^{2}}{1+(\mu_{h}B)^2}-\frac{n_{e}\mu_{e}^{2}}{1+(\mu_{e}B)^2}\right],
\end{gather}
where $n_{e}$, $n_{h}$, $\mu_{e}$, and $\mu_{h}$ are electron-type
carrier density, hole-type carrier density, electron-type mobility, 
and hole-type mobility, respectively. Both the electron and hole
mobility decrease monotonously as temperature increases, as seen
in Fig. 5(f), which is a typical behavior for metals. Intriguingly,
the electron-hole compensation only holds at low temperatures and
becomes ineffective above 50 K. The Hall coefficient obtained by
fitting with the two-band model
$R_{H}=[R_{H}^{h}(\sigma_{xx}^{h})^{2}-R_{H}^{e}(\sigma_{xx}^{e})^{2}]/(\sigma_{xx}^{h}+\sigma_{xx}^{e})^{2}$ \cite{twobandhall_RAF_PRL2009}, 
as shown in Fig. 5(e), is quite consistent with the directly
measured $R_{H}$ which is acquired from the slope of $\rho_{yx}$
near the zero field, where $R_{H}^{h}$ and $R_{H}^{e}$ are the Hall
coefficient for hole-type and electron-type charge carriers
respectively, $\sigma_{xx}^{h}$ and $\sigma_{xx}^{e}$ are the hole
conductivity and electron conductivity, respectively, and
$\sigma_{xx}=en\mu$. All of above fitting results suggest that the
two-band model fits the experimental data reliably. It is apparent
that the absolute value of $R_{H}$ changes little below 14 K and becomes large with
increasing temperature, then becomes small again above 100 K, at
which temperature the slope of $\rho_{yx}(B)$ reaches the maximum
[Fig. 5(b)]. We propose that the shift of Fermi level with
increasing temperature may account for the loss of the
electron-hole compensation above 50 K. Similar behavior and
explanations were reported in WTe$_{2}$ \cite{WTe2_WuY_PRL15} and
ZrTe$_{5}$ \cite{ZrTe5Lifshitz_ZhangY_natcom17}.

In summary, we have discovered a topological semimetal
W$_{2}$As$_{3}$ with $Z_{2}$ indices [1;111] which has a strong TI
feature. The electron-hole compensation leads to a colossal MR
which is as large as 11 500\% at 3 K and 15 T and is unsaturated
even up to 60 T. A nontrivial Berry's phase of $\pi$ is obtained
from $F_{h3}$, $F_{h4}$, and $F_{h5}$, suggesting the nontrivial topological
characteristic. A two-band model is effective to fit the Hall
conductivity and Hall coefficient. The transport properties are in
good agreement with the band structure calculations.

\section*{acknowledgments}

We thank Yi Zhou, Yongkang Luo, and Chenqiang Hua for insightful
discussions. This work was supported by the National Key R\&D
Program of the China (Grants No. 2016YFA0300402 and No. 2014CB648400),
the National Science Foundation of China (Grants No. 11774305 and No. 
11274006), and the Fundamental Research Funds for the Central
Universities of China.

%

\end{document}